# Frequency-dependent Ultrasonic Stimulation of Poly(N-isopropylacrylamide) Microgels in Water

Atieh Razavi[1], Matthias Rutsch[2], Sonja Wismath[2], Mario Kupnik[2], Regine von Klitzing[1], and Amin Rahimzadeh[1,*]

[1] Soft Matter at Interfaces, Institute for Condensed Matter Physics, Technische Universität Darmstadt, Hochschulstraße 8, 64289 Darmstadt, Germany.
[2] Measurement and Sensor Technology, Department of Electrical Engineering and Information Technology, Technische Universität Darmstadt, Merckstraße 25, 64283 Darmstadt, Germany.
* Correspondence: amin.rahimzadeh@pkm.tu-darmstadt.de;

**Abstract:** As a novel stimulus, we use high-frequency ultrasonic waves to provide the required energy for breaking hydrogen bonds between Poly(N-isopropylacrylamide) (PNIPAM) and water molecules while the solution temperature maintains below the volume phase transition temperature (VPTT=32 °C). Ultrasonic waves propagate through the solution and their energy will be absorbed due to the liquid viscosity. The absorbed energy partially leads to the generation of a streaming flow and the rest will be spent to break the hydrogen bonds. Therefore, the microgels collapse and become insoluble in water and agglomerate, resulting in solution turbidity. We use turbidity to quantify the ultrasound energy absorption and showed that the acousto-response of PNIPAM microgels is a temporal phenomenon that depends on the duration of the actuation. Increasing the solution concentration leads to a faster turbidity evolution. Furthermore, an increase in ultrasound frequency leads to an increase in the breakage of more hydrogen bonds within a certain time and thus faster turbidity evolution. This is due to the increase in ultrasound energy absorption by liquids at higher frequencies.

**Keywords:** Poly(N-isopropylacrylamide); Microgels; Ultrasound; Turbidity; Hydrogen bond; Acousto-responsive

## 1. Introduction

Microgels are cross-linked polymeric networks, in a size range of nanometer to micrometer. These particles swell in water and according to their environment, their size and density can be adjusted. They are stimuli-responsive due to their ability to undergo a volume phase transition (VPT) subjected to the changes in temperature, pH, ionic strength, and so on [1]. These smart polymers have drawn attention for several years since they have promising applications in a wide range of industries from biomedical technologies, cement, ink-jet printing, oil recovery, etc. [2].

Poly(N-isopropylacrylamide) –known as PNIPAM – is a thermoresponsive microgel having a volume phase transition temperature (VPTT) of 32 °C [3], [4]. At room temperature (below VPTT), PNIPAM is fully swollen and surrounded by hydrogen-bonded water molecules– approximately 4-6 hydrogen bonds per NIPAM monomer are formed [5]. By increasing their solution temperature above the VPTT, their hydrogen bonds break, and microgels collapse and become insoluble in water. The required thermal energy for breaking the hydrogen bonds is $2K_BT$ per NIPAM monomer [6].

PNIPAM microgels have been designed to be responsive to other stimuli such as pH [7,8], light [9], magnetic field [10], ionic strength [11], [12], or a combination of these parameters [11,13, 14] in addition to their natural responsiveness to changes in temperature (thermal energy). PNIPAM-based pH-sensitive microgels have been synthesized by copolymerization of acrylic acid (AAc) with PNIPAM [7]. In addition to pH, these microgels show sensitivity to changing the electrolyte concentration by increasing the amount of AAc [11]. Furthermore, the sorption of a photosensitive surfactant by PNIPAM microgels leads to the formation of light-sensitive microgels that swell and de-swell by irradiation of UV light and visible light, respectively [9].

In all of these cases, one has to modify the PNIPAM microgel by copolymerization so it becomes responsive to another stimulus. Recently, we have shown that PNIPAM microgels, without any additional modification, are acousto-responsive [15]. Hydrogen bonds between PNIPAM and water molecules will be broken due to the absorption of a certain amount of energy carried by ultrasound waves while the solution temperature is below VPTT. Sound absorption in liquids has been the subject of research for a long time [16]. The intensity of the ultrasound at a distance ($x$) from the input wave source can be calculated by $I = I_0 \exp(-2\alpha x)$. Where $I_0$ denotes the intensity at $x = 0$ and $\alpha$ is the wave attenuation/absorption coefficient. Hundreds of years of research reveal that the absorption coefficient is proportional to the liquid viscosity due to the relaxation of translational, vibrational, and rotational degrees of molecular freedom [17]. If the molecules, that are excited vibrationally, return to the equilibrium state in a shorter time than the exciting ultrasound period, the vibrational energy of molecules returns to the control volume in the same phase as the input waves. As a result, the net energy loss due to vibrational modes will be zero. As the frequency of the actuating ultrasound increases, the energy of vibrating molecules returns out of phase with the input ultrasound period and will be detected as energy loss. Finally, by increasing the wave frequency to a large enough value, there will be not enough time for the molecules to exchange their energy between translational and vibrational modes. In the case that the actuating ultrasonic frequency is comparable to the inverse time required for the energy exchange, the input frequency is called relaxation frequency [18]. The aforementioned theory is important in the analysis of hydrogen bond breakage and turbidity evolution due to the absorption of ultrasound energy at different frequencies.

In this paper, we further elaborate on the acousto-responsiveness of PNIPAM microgels by using the turbidity evolution of PNIPAM solutions subjected to ultrasonic actuation. We measure the ultrasound energy loss in the solutions corresponding to the required energy for breaking hydrogen bonds between PNIPAM microgels and water molecules. As a result, the solution becomes turbid due to the agglomeration of the collapsed microgels. In our recent work [15], we showed that PNIPAM microgels are responsive to ultrasound energy, and by increasing the input power, turbidity evolution becomes faster. The question remains on the effects of input frequency on turbidity evolution (i.e., hydrogen bond breakage) of PNIPAM solutions. Therefore, in the present study, the acousto-responsiveness of the PNIPAM microgels is introduced as a function of ultrasound frequency by keeping the input acoustic pressure constant. We used turbidity to quantify the ultrasound energy absorption by solutions, using image processing.

## 2. Results and Discussion

*2.1. Turbidity evolution*

Ultrasonic waves created by the piezoelectric transducer propagate into the PNIPAM solution and by dissipation of their energy (due to the liquid viscosity) into the solution the energy that is required to break the hydrogen bonds will be provided. Therefore, the PNIPAM polymers collapse and become insoluble in water, the solution becomes turbid and the agglomeration of polymers can be visualized by a camera. Continuous actuation of the liquid results in breaking more hydrogen bonds, and, thus, more turbidity arises until the entire medium becomes fully white. In order to break the hydrogen bonds with thermal energy, one needs to provide 5 kJ per mol of NIPAM in the water [19]. Therefore, according to our experiments, approximately 2.2 J, 0.44 J and 0.088 J are required to break all hydrogen bonds inside 1 ml of 5 wt%, 1 wt% and 0.2 wt% solutions, respectively.

The turbidity evolution of 1 wt% microgel solution is shown sequentially for 2.5 MHz and 30 mV (150 kPa sound pressure inside the cuvette) of input actuation in Figure 1. In order to break the hydrogen bonds of PNIPAM-water, and thus turbid the medium, the calculated amount of energy is needed as discussed above. This energy is provided by ultrasound waves instead of thermal energy and equals the absorbed energy of the acoustic waves by the liquid and not the total energy that waves have. Continuing the actuation for a certain time collects the absorbed energy from the acoustic field gradually and the expected stored energy for appearing turbidity will be provided. Turbidity starts from the bottom and extends to the whole solution in less than 90 seconds.

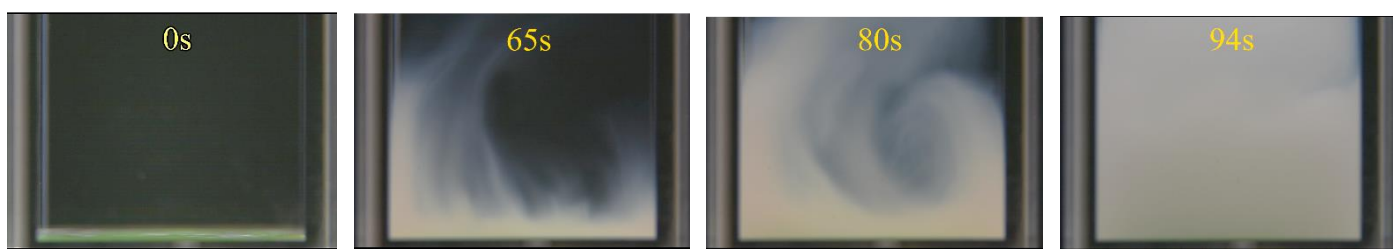

**Figure 1.** Acoustic wave energy absorption (turbidity) mechanism during the imposing 2.5 MHz and 30 mV ultrasonic actuation of the 1 wt.% microgel solution.

## 2.2. Effects of solution concentration

The kinetics of the energy absorption has been quantified similarly to our previous work [15] using image processing and the absorbed energy over time for different solution concentrations (i.e., 0.2, 1, and 5wt%) is shown in Figure 2. Increasing the solution concentration increases the ultrasound energy absorption and consequently increases the speed of turbidity evolution. This is because in more dense solutions, the PNIPAM chains are slower in rearranging their structure upon ultrasonic actuation and therefore the energy attenuation is larger.

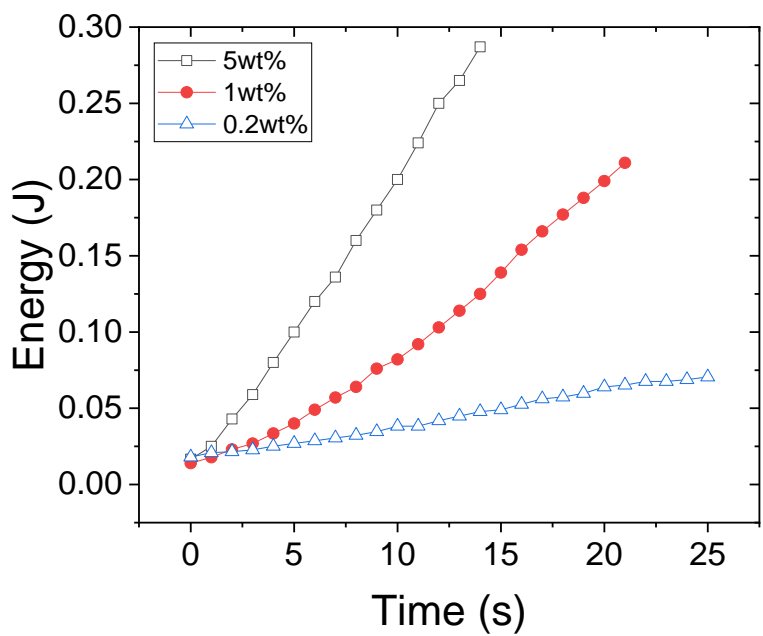

**Figure 2.** The energy associated with the broken hydrogen bonds due to 2.5 MHz ultrasound waves in solution concentrations of 0.2, 1, and 5 wt%.

*2.3. Effects of frequency of ultrasound*

In addition to solution concentration, the effect of ultrasound frequency was investigated at a fixed input ultrasound pressure. The absorbed ultrasound energy in 1 wt% PNIPAM solution versus time due to the actuation frequency, ranging from 40 kHz to 5 MHz, is shown in Figure 3. According to Figure 3, low-frequency actuation leads to lower energy absorption and therefore a longer turbidity evolution, ultimately a frequency exists (i.e., 40 kHz) in which the turbidity does not occur even after 10 minutes of actuation. This is because at the 40 kHz actuation the liquid does not absorb enough ultrasound energy and the absorbed energy is spent solely on acoustic streaming. Furthermore, the streaming flow helps the microgels to regain their bonding strength during the mixing. In contrast, when the ultrasound frequency is in the MHz range, the entire solution becomes turbid in less than two minutes. The total absorbed energy divided by the required time to break all hydrogen bonds within the solution is calculated for all frequency cases to define a new parameter as 'absorbed power' for different frequencies. The result (Figure 4) is consistent with the theory and experiments stating that liquids absorb more power from ultrasound waves that have higher frequencies [20]. From Figure 3, one can infer that the relaxation frequency for the fixed input ultrasound power falls between 40 kHz and 260 kHz, where the hydrogen bonds between PNIPAM and water can regain their strength in the period of the input ultrasound waves.

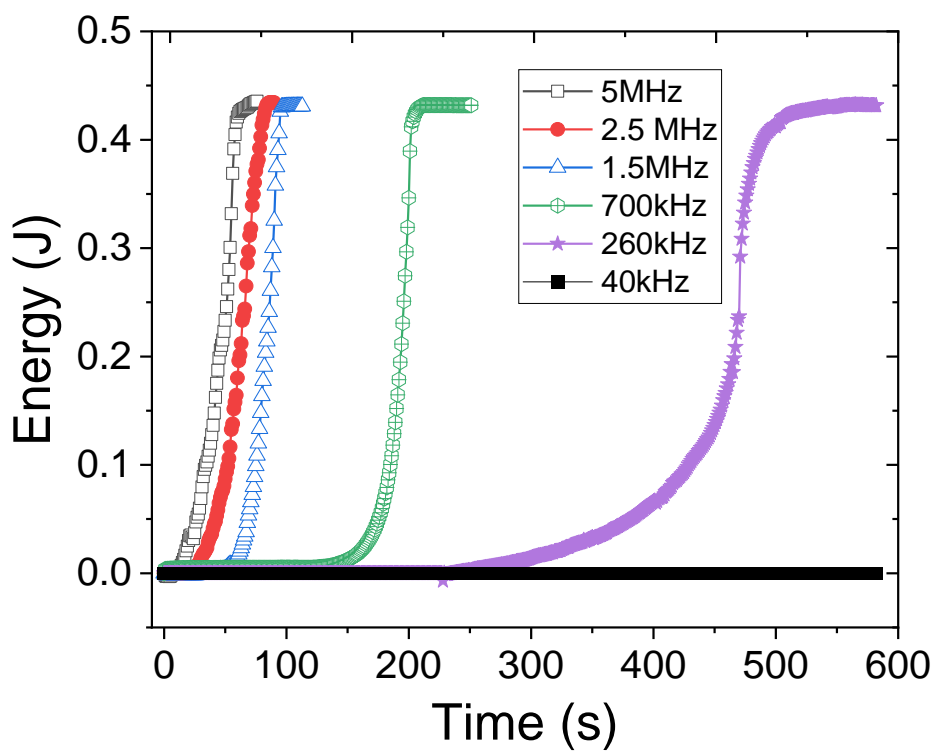

**Figure 3.** Ultrasound energy absorption during actuation of 1 wt% PNIPAM solution in different frequencies.

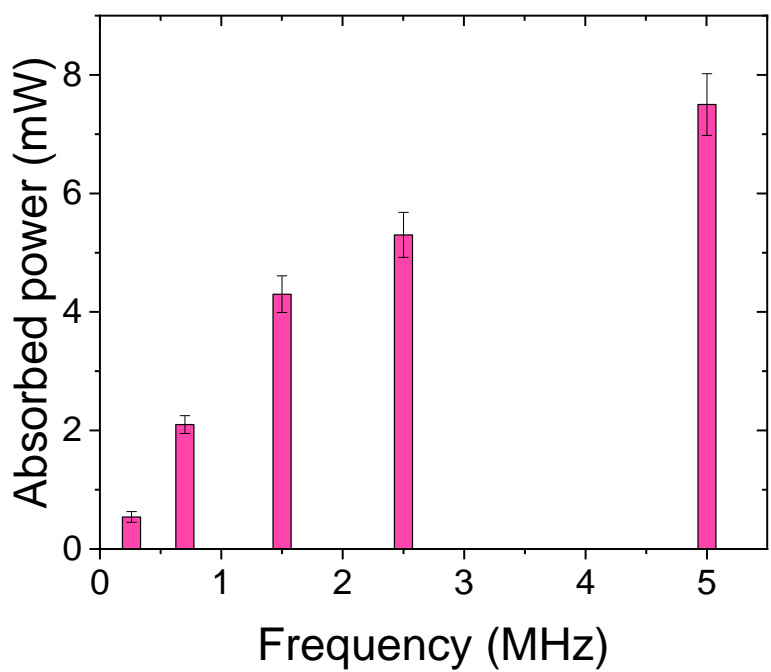

**Figure 4.** Acoustic power absorption by 1 wt% PNIPAM solution subjected to ultrasound waves in different frequencies.

According to the aforementioned literature [21], imposing ultrasound waves on a liquid generates translational movement of the liquid molecules which is called acoustic streaming. The question is how the turbidity evolution can be related to this acoustic streaming. Do all hydrogen bonds break at the bottom of the glass cuvette and collapsed microgels will be transferred to the rest of the cuvette due to the streaming? In order to address this question, we added polystyrene particles to the solutions to visualize the particle trajectory. If the particle trajectory is the same as the turbidity evolution pattern, we can conclude that the microgels collapse at the bottom and travel to the rest of the media. Figure 5 shows the particle trajectory and hence the streaming pattern due to the actuation of 1 wt% PNIPAM solution subjected to 2.5 MHz ultrasound waves. The streaming pattern in Figure 5 shows a clear distinction from the turbidity evolution pattern shown in Figure 1. Therefore, we can conclude that hydrogen bonds break not only at the bottom of the cuvette but wherever the required energy for breaking them is provided by the complex acoustic field due to the reflection of waves on the walls of the cuvette. On the other hand, we cannot claim that the turbidity evolution is only due to the breaking of hydrogen bonds locally since the results, shown in Figure 5, proves the existence of the streaming inside the solution. Therefore, the turbidity evolution pattern is complex and consists of both streaming flow and hydrogen bond breakage in the entire solution.

By exposing a PNIPAM solution to ultrasonic waves, the liquid absorbs enough energy to start the translational movement of molecules, and the additional absorbed energy will be spent to break the hydrogen bonds. In addition, all of the hydrogen bond breakages cannot be visualized since microgel agglomeration can only be observed (via turbidity). Therefore, the averaged absorbed energy over time is a more suitable term to compare in different cases (Figure 4).

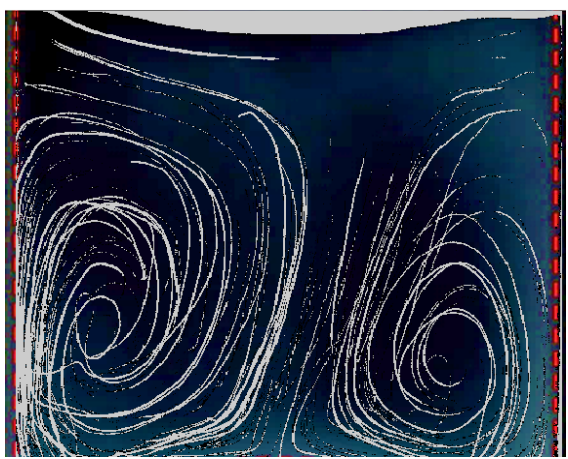

**Figure 5**. The streaming pattern inside the 1 wt% PNIPAM solution due to 2.5 MHz ultrasound waves, visualized by the polystyrene particle's trajectory.

## 3. Conclusion

In this paper, we showed that PNIPAM microgels are responsive to ultrasonic actuation and this acousto-response depends on the duration of the actuation in contrast to the thermo-response of PNIPAM microgels. When the solution is subjected to ultrasonic waves, the ultrasound energy will be absorbed by the solution due to its viscosity resulting in the translational, vibrational and rotational relaxation of liquid molecules. The absorbed energy due to the translational relaxation leads to acoustic streaming, and the rest of the absorbed energy provides the required energy for breaking hydrogen bonds between PNIPAM and water molecules. As a result, microgel particles become insoluble in water, agglomerate and can be seen as a turbid medium. Therefore, no matter how much energy the input waves have, the absorbed energy by the solution accumulates ultimately during a certain time, and if the collected absorbed energy reaches a threshold it starts to break the hydrogen bonds. However, this may not lead to hydrogen bond breakage and turbidity in low frequencies, since the absorbed energy is not enough and the mixing of solution due to the acoustic streaming helps microgels to regain their hydrogen-bond strength. We used this phenomenon to quantify the ultrasonic energy absorption in PNIPAM solutions. We showed that the solution concentration decreases the time for turbidity evolution since solutions with higher concentrations absorb more ultrasound power. Furthermore, by increasing the ultrasound frequency the solution absorbs more power and turbidity arises faster. Therefore, one can conclude that ultrasound waves having a higher frequency can be a faster stimulus for PNIPAM microgels.

## 4. Materials and Methods

Linear Poly(N-isopropylacrylamide) (PNIPAM) ($M_n \approx 40,000 \; g/mol$)) was purchased from Sigma-Aldrich, Germany. Three different concentrations of 0.2 wt%, 1 wt%, and 5 wt% were prepared by dissolving it in Milli-Q water and stirred over one night. A fresh microgel solution (1 ml) of each sample was poured inside a Quartz SUPRASIL Macro Cell (PerkinElmer, MA, USA) cuvette (inside dimensions of $10 \times 10 \times 40 \; mm^3$) and attached to the ultrasonic transducer surface using a two-component latex glue. Piezo-ceramic ultrasonic transducers (STEMINC-PIEZO, Davenport, USA) with a resonance frequency of 5 MHz, 2.5 MHz, 1.5 MHz, 700 kHz, 260 kHz, and 40 kHz were used to transfer acoustic waves into the liquid (Figure 1). A 3D-printed holder was designed and fabricated to hold the transducer with the attached cuvette. A function generator (SDG1062X, SIGLENT, China) was used to generate RF signals. The signal was amplified by an RF amplifier (VBA100-30, Vectawave, UK) and connected to the transducer. Input voltages were carefully selected to create the same vibration amplitude (or sound pressure) in all frequencies. The sound pressure was measured using a needle-shaped hydrophone (HNR-1000, Onda, USA) at the bottom of the cuvette and was adjusted at $150 \pm 20$ kPa in all frequencies. The experiments were carried out at an ambient temperature of around 22 °C. Continuous actuation of transducers does not increase their temperature by more than 3 °C, making sure the PNIPAM solutions maintain below their VPTT =32 °C. A macro lens (TAMRON, SP 90mm, F/2.8) mounted on a Nikon D7200 camera was used to visualize the turbidity in the solutions by 60 frames per second (FPS) videos (Figure 6).

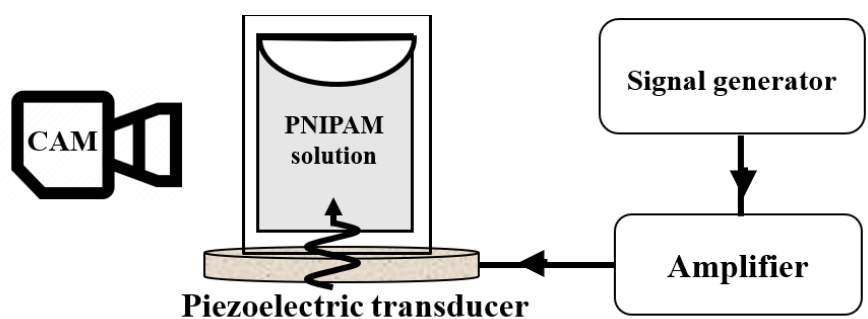

**Figure 6.** The experimental setup consists of the electrical section (function generator, oscilloscope and amplifier), the optical section (camera), and the sample.

**Author Contributions:** Conceptualization, At.R., A.R. and R.v.K.; methodology, At.R. and A.R.; formal analysis, At.R.; investigation, At.R., M.R., and S.W.; resources, M.R., S.W. and M.K.; writing-original draft preparation, At.R. and A.R.; writing-review and editing, At.R., M.R., S.W., M.K., R.v.K., and A.R.; supervision, M.K. and R.v.K. and A.R.; project administration, M.K. and R.v.K.; funding acquisition, A.R. and R.v.K. All authors read and agreed to the published version of the manuscript.

**Funding:** This research was funded by the German Research Foundation (DFG) - Project Number (460540240).

**Data Availability Statement:** Not applicable.

**Acknowledgments:** We acknowledge support by the German Research Foundation (DFG).

**Conflicts of Interest:** The authors declare no conflict of interest.